\documentclass[aps,preprint]{revtex4}
\usepackage{amsmath}
\usepackage{amssymb}
\usepackage{mathrsfs}
\usepackage{xcolor}
\usepackage{graphicx}
\usepackage{subfigure}
\usepackage{enumerate}
\usepackage{float}
\usepackage{booktabs}
\usepackage{diagbox} 
\usepackage{array}
\usepackage{tabularx}
\begin{document}
	\title{Stationary and Free-fall frame Kerr black hole in gravity's rainbow}	 
	\author{Yuzhou Tao$^{a}$}
	\email{taoyuzhou@stu.scu.edu.cn}
	\author{Benrong Mu$^{a}$$^{,}$$^{b}$}
	\email{benrongmu@cdutcm.edu.cn}
	\author{Siyuan Hui$^{a}$}
	\email{huisiyuan@stu.scu.edu.cn}
	\author{Jun Tao$^{a}$}
	\email{taojun@scu.edu.cn}
	\affiliation{$^{a}$Center for Theoretical Physics, College of Physical Science and Technology,
		Sichuan University, No. 24 South Section 1 Yihuan Road, Chengdu, China}
	\affiliation{$^{b}$Physics Teaching and Research section, College of Medical Technology,
	Chengdu University of Traditional Chinese Medicine,
	No. 1166 Liutai Avenue, Chengdu, China}
	\begin{abstract}
		Doubly special relativity (DSR) is an effective model for encoding quantum gravity in flat spacetime. To
		incorporate DSR into general relativity, one could use “gravity’s rainbow,” where the spacetime
		background felt by a test particle would depend on its energy. In this paper, we investigate the thermodynamics of rainbow Kerr black hole in the scenario with the stationary(ST) orthonormal frame and free-fall(FF) orthonormal frame. After the rainbow metric in ST frame and FF frame is deduced, the Hamilton-Jacobi method is used to acquire the modified Hawking temperature, specific heat and corresponding the modified entropy to each scenario, then the thermodynamic properties are discussed. We find that the effects of rainbow gravity on Kerr black holes are quite model-dependent. In other words, the value of parameter $\eta$ and $n$  with Amelino-Camelia's proposal are crucially important and worth discussing. Specificly, with most widly accepted choice ($n=2,\eta >0$), the effects of rainbow gravity tend to decrease the Hawking temperature but increase the black hole entropy in ST frame, and increase the Hawking temperature but decrease the black hole entropy in FF frame conversely.
	\end{abstract}
    \maketitle
	\tableofcontents	
  	\section{Introduction}
     The study of black hole thermodynamics has been playing
     an increasingly prominent role in our understanding of the
     interdisciplinary area of general relativity, quantum mechanics, information theory and statistical physics. Since Hawking studied the thermodynamics of black holes by combining general relativity with quantum field theory\cite{Hawking:1974rv,Hawking:1975vcx}, the relationship between entropy and event horizon area has been successfully developed \cite{Bekenstein:1973ur,Bekenstein:1974ax}. After this discovery, people realized that there is a Trans-Planckian problem in Hawking's work\cite{Unruh:1976db} and that Hawking's prediction depends on the validity of quantum field theory to arbitrary high energy in curved
     spacetime. Moreover, 
     quantum field theory reveals that Lorentz symmetries can be modified at high energies. 
      
     A common feature of some quantum gravity theories is the modification of the dispersion relation, namely "Double special relativity"(DSR),
     which considers the speed of light and the Planck energy scale as two constants of nature, and guarantees nonlinear Lorentz transformations in momentum spacetime \cite{Amelino-Camelia:2000cpa,Amelino-Camelia:2000stu,Magueijo:2001cr,Magueijo:2002am}. MDR might play an important
     role in astronomical and cosmological observations, such as the threshold anomalies of ultra high energy cosmic rays and TeV photons \cite{Amelino-Camelia:1997ieq,Colladay:1998fq,Coleman:1998ti,Amelino-Camelia:2000bxx,Jacobson:2001tu,Jacobson:2003bn}, and ground observations and
     astrophysical experiments have tested the predictions of MDR theory \cite{Mittleman:1999it,Cane:2003wp,Shao:2011uc,Petry:1999fm}.
     To incorporate DSR into the framework of general
     relativity, Magueijo and Smolin proposed the “gravity’s rainbow,” where the spacetime background felt by a
     test particle would depend on its energy \cite{Amelino-Camelia:2005zpp}.
     Specifically, the modified energy-momentum dispersion relation of particle with energy $E$ and momentum $p$ takes the following form 
     \begin{equation}\label{1}
     	E^{2}f^{2}(\frac{E}{E_{p}})-p^{2}g^2(\frac{E}{E_{p}})=m^2,
\underline{\underline{}}     
\end{equation}
     where $E_{p}$ is the Planck energy.
     The two functions called rainbow function $f(x)$ and $g(x)$ should have following properties
     \begin{equation}
     	\lim\limits_{x\to0}f(x)=1,\qquad \lim\limits_{x\to0}g(x)=1.
     \end{equation}
     Amelino-Camelia proposed a popular choice for functions $f(x)$ and $g(x)$\cite{Amelino-Camelia:1996bln,Amelino-Camelia:2008aez}
     \begin{align}\label{2}
     	f(x)=1,\qquad g(x)=\sqrt{1-\eta x^n}.
     \end{align}
     Since then, it has
     been extensively studied to explore various aspects of black holes and cosmology\cite{Galan:2004st,Hackett:2005mb,Aloisio:2005qt,Ling:2005bp,Garattini:2011hy,Garattini:2011fs,Amelino-Camelia:2013wha,Barrow:2013gia,Garattini:2014rwa,Mu:2015qna,Ali:2014zea,Gangopadhyay:2016rpl,Gim:2015yxa,Kim:2016qtp,MahdavianYekta:2019dwf,Mu:2019jjw}.
     eq:(\ref{2}) is compatible with some results obtained in the loop quantum gravity method, and we dub $n$ and $\eta$ as exponential parameter and effect parameter respectively. The phenomenological meaning of  "Amelino-Camelia (AC) dispersion relation" is also reviewed \cite{Amelino-Camelia:2008aez}.
     Combined with equation(\ref{1}), equation(\ref{2}) and Heisenberg uncertainty principle, we can express $E$ in defferent
     case where we let $n=0,1,2,4$ in TABLE \ref{20}. The usual dispersion
     relation is a quadratic polynomial and gives us two values for the energy, one positive and
     another negative. For  $n>2$, there could be more than two solutions, but they aren't all physically acceptable.
     
     As regards the metric, it would be replaced by a one-parameter family of metrics which depends on the energy of the test particle, 
     forming a “rainbow metric”. Specifically, for a Kerr black hole, the 
     corresponding “rainbow metric” solution in a stationary orthonormal frame is given.
     Since the rainbow metric is the metric that the radiated particles 
     “see”, a more natural orthonormal frame is the one anchored to 
     the particles. Therefore, the rainbow 
     Kerr black hole in the free-fall orthonormal frame is worth discussing. Many important and groundbreaking work on free-fall frame was carried out in Ref.\cite{Gim:2015yxa}.
         \begin{table}[t]\label{20}
     	\centering
     	\caption{the solved energy of defferent $n$}
     	
     	\begin{tabular}{|p{3cm}|p{4cm}|p{4cm}|}
     		\hline
     		the choice of $n$&solution 1 of $E$&solution 2 of $E$   \\
     		\hline
     		$n=0$& $\sqrt{m^2+\frac{1-\eta}{r_{+}^2}}$&none     \\
     		\hline
     		$n=1$& $\frac{-\eta + \sqrt{4m^2r_{+}^4+4r_{+}^2+\eta^2}}{2r_{+}^2}$ &  $\frac{-\eta- \sqrt{4m^2r_{+}^4+4r_{+}^2+\eta^2}}{2r_{+}^2}$    \\
     		\hline
     		$n=2$&$\sqrt{\frac{1+ m^2 r_{+}^2}{r_{+}^2+\eta}}$&none     \\
     		\hline
     		$n=4$&	$\sqrt{\frac{-r_{+}^4+\sqrt{r_{+}^4+4\eta(m^2r_{+}^2+1)}}{2\eta}}$& $\sqrt{\frac{-r_{+}^4-\sqrt{r_{+}^4+4\eta(m^2r_{+}^2+1)}}{2\eta}}$       \\
     		\hline 
     	\end{tabular}
     \end{table}

     \begin{figure}[t]\label{fig1}
     	\centering
     	\subfigure[the solution of energy with $\eta>0$.]{
     		\includegraphics[scale=0.6]{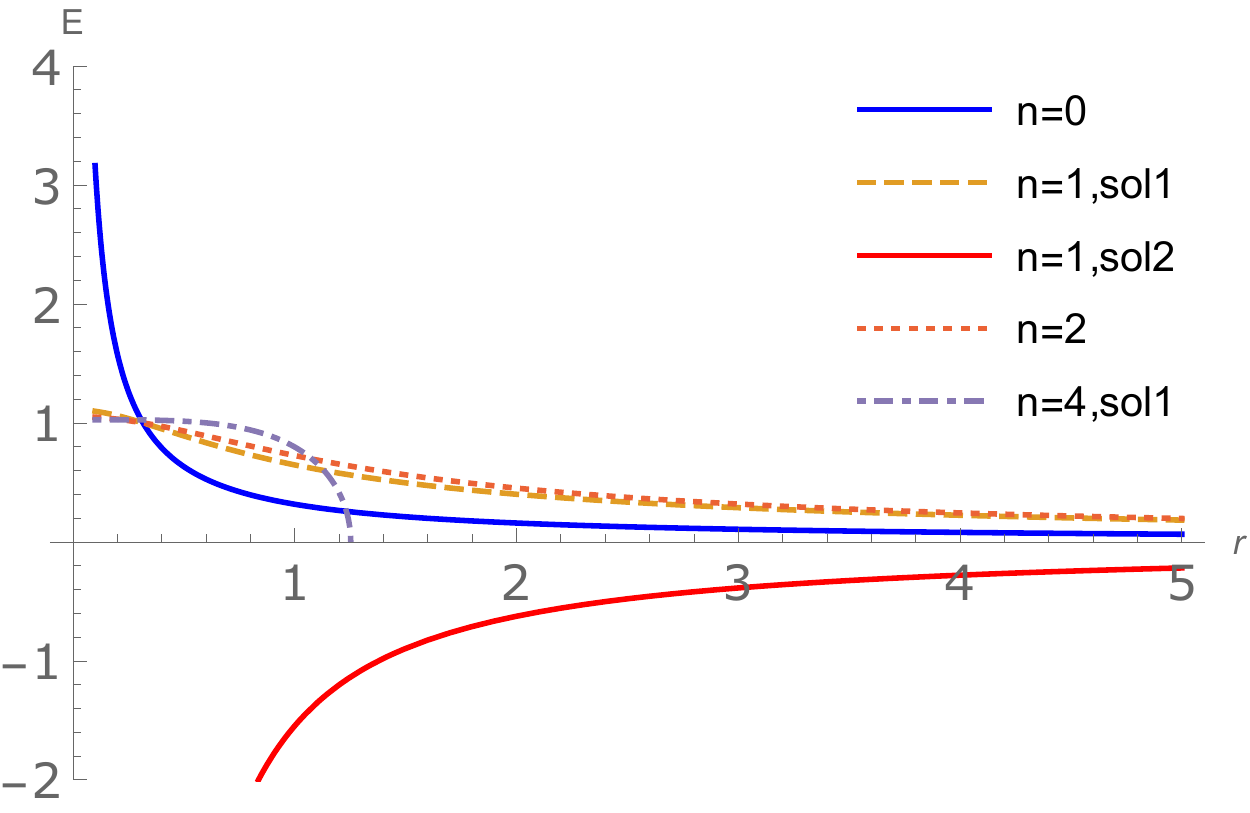}}
     	\quad
     	\subfigure[the solution of energy with $\eta<0$.]{
     		\includegraphics[scale=0.6]{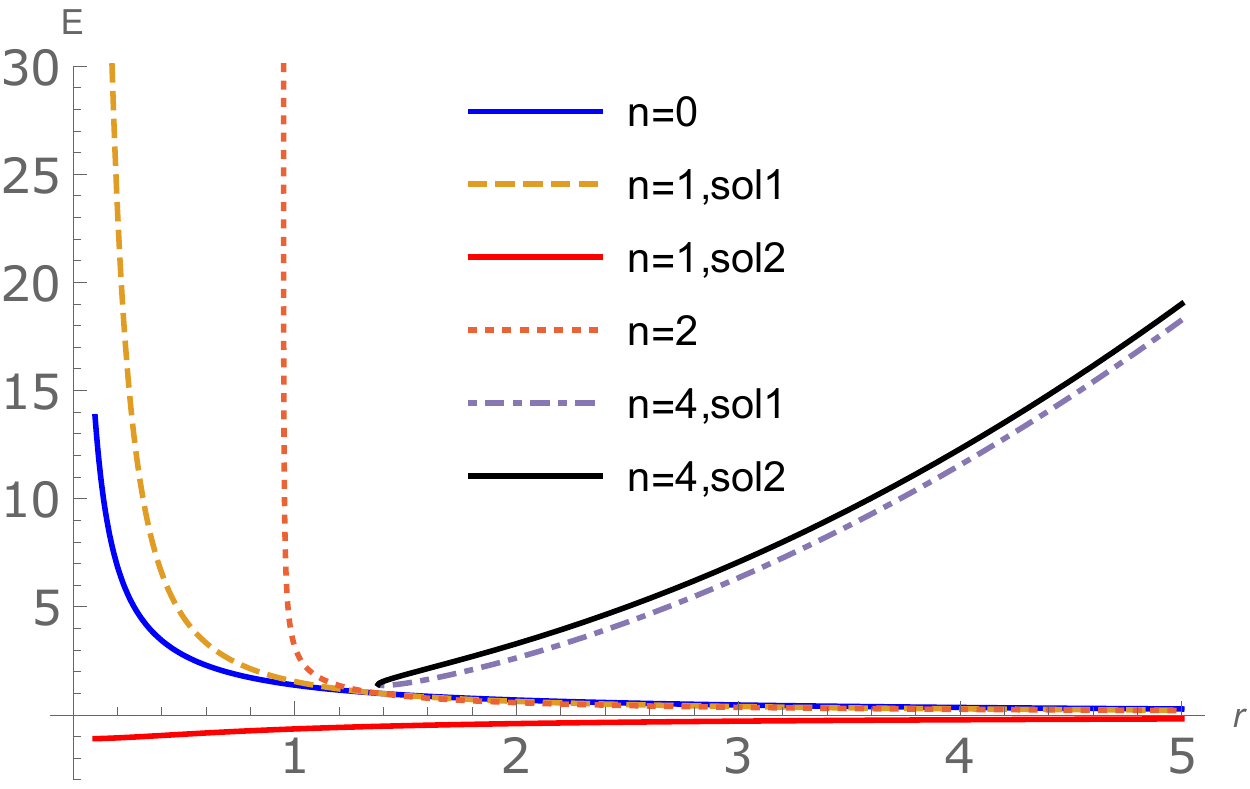}}
     	\caption{The probed particle's energy $E(r)$ expressed by horizon radius $r_{+}$ with various values of $n$. }
     \end{figure}

     Recently, a semi-classical method to simulate Hawking radiation as a tunneling process has been developed. This method was first proposed by Kraus and Wilczek and is called zero-geodesic method\cite{Kraus:1994by,Kraus:1994fj}. Subsequently, the Hamilton-Jacobi method is used to study the tunneling of particles\cite{Srinivasan:1998ty,Angheben:2005rm,Kerner:2006vu},which has become the mainstream method to calculate the thermodynamics of black holes since then, and the mainstream correction mainly includes tunnel effect correction and rainbow gravity correction.  
     Hawking radiation of black holes based on particles in a dynamical geometry is studied\cite{Parikh:1999mf}. Then the tunneling effect is studied by using the semiclassical method \cite{Kerner:2007rr,Kerner:2008qv}. In addition, the Hamilton-Jacobi equation is modified considering the influence of quantum gravity, and the modified Hawking temperature is derived\cite{Chen:2013pra,Chen:2013tha,Chen:2013ssa,Chen:2014xsa,Chen:2014xgj,Mu:2015qta}, which inspire us to study the gravitational rainbow effect of Hawking radiation \cite{Mu:2015qna,Tao:2016baz}.

    In the remainder of our paper, The two rainbow Kerr black holes are labelled
    by stationary frame (ST) and free-fall frame (FF) rainbow Kerr black holes with defferent metrics respectively. In section \uppercase\expandafter{\romannumeral2},  we will present the modified rainbow Kerr metrics in ST frame, as well as we will discuss black hole thermodynamics in special cases in ST frame. In section \uppercase\expandafter{\romannumeral3}, the metric of a free-fall rainbow Kerr black hole is derived, then its thermodynamics are obtained and disscussed. Finally, section \uppercase\expandafter{\romannumeral4} is devoted to our discussion and conclusion. Throughout the paper we take geometrized units $c = G = k_{b} = 1$, where the Planck
    constant $\hbar$ is square of the Planck mass $m_{p}$.
	        
	\section{Thermodynamics in stationary frame}
	 The Kerr black hole is an important solution to Einstein field
	 equation in a $(3 + 1)$ dimensional space, whose metric is\cite{Kerr:1963ud}
	 \begin{equation}
	 	ds^2=-(1-\frac{2Mr}{\rho})dt^2-\frac{4aMr\sin^2 {\theta}}{\rho}dtd\phi+\frac{\rho}{\Delta}dr^2+\rho d\theta^2+(\Delta+\frac{2Mr(r^2+a^2)}{\rho})	\sin^2\theta d\phi^2,
	 \end{equation}
	 where 
	 \begin{equation}
	 	\rho=r^2+a^2\cos^2\theta,
	 \end{equation}
	 and
	 \begin{equation}
	 	\Delta=r^2-2Mr+a^2.
	 \end{equation}
	 Here $a=J/M$ is the spin parameter, which means angular momentum per unit mass. 
	 $\Delta$ vanishes when $r_{\pm}=M\pm\sqrt{M^2-a^2}$, while $g_{tt}$ vanishes when $r=r_{S\pm}=M\pm\sqrt{M^2-a^2\cos^2\theta}$.
	With rainbow gravity, the metric becomes
	\begin{equation}
	\label{eq:5}
	\begin{aligned}
	ds^2(E)=&-\frac{(1-\frac{2Mr}{\rho})}{f^2(x)}dt^2-\frac{4aMr\sin^2 {\theta}}{\rho f(x)g(x)}dtd\phi+\frac{\rho}{\Delta g^2(x)}dr^2 \\ 
	&+\frac{\rho}{g^2(x)} d\theta^2+\frac{(\Delta+\frac{2Mr(r^2+a^2)}{\rho})sin^2\theta}{g^2(x)} d\phi^2.
	\end{aligned}
	\end{equation}
	
     In the background of rainbow Kerr metric, the action of a tested particle is
	\begin{equation}
	I=A(t)+B(\phi)+W(r)+\Theta(\theta)+\zeta.
	\end{equation}
	 As the angular momentum, energy, and Hamiltonian of particle are conserved, the ST rainbow black hole is independent of t and $\phi$.
	 Separate the variables from the action, we can employ the following ansatz for the action $I$
	\begin{equation}
		\partial_{t}I=\omega,\qquad
		\partial_{\phi}I=j.
	\end{equation}
	Combine the Hamilton-Jacobi equation and metric (\ref{eq:5}), we get
	\begin{align}\label{7}
	Pr^2&=g_{rr}(2g^{t\phi}\omega j-g^{\theta\theta}P_{\theta}^2-g^{\phi\phi} j^2-g^{tt} \omega^2-m^2).
	\end{align}	
	The $Pr$ is the momentum in radial direction. By using the residue theory, we get\cite{Mu:2019jjw}
	\begin{equation}\label{16}
	ImI_{\pm}=\int Pr_{\pm}dr.
	\end{equation}
	The $\pm$ denotes outgoing/ingoing solutions, and $ImI_{\pm}$ means the imaginary part of action of probed particle.
	
	For a particle of energy $E$ and angular momentum $L$ residing in a
	system with temperature $T$ and angular velocity $\omega$, the Maxwell–
	Boltzmann distribution is\cite{Wang:2015zpa,Tao:2015dhe}
	\begin{equation}
	P \propto \exp{(-\frac{E-\omega L}{T})},
	\end{equation}
	and the articles give the probability of a particle tunneling from inside to outside of
	the horizon, where effective Hawking temperature can be read off
	from \cite{Hawking:1975vcx,Unruh:1976db}
	\begin{equation}\label{9}
	T=\frac{\hbar\omega}{2(ImI_{+}-ImI_{-})}.
	\end{equation}
    From equation(\ref{7}), we find 
	\begin{equation}
	Pr_{\pm}=\pm \frac{1}{\Delta g}\sqrt{(-Ma\omega g jr +a^2j^2+\omega ^2(r^2+a^2)^2)-\Delta (g^2 P_{\theta}^2+\rho^2m^2+\frac{j^2}{\sin^2 \theta}+a^2\sin^2\theta\omega^2)}.
	\end{equation}
	To numerically investigate $T(M)$ and $S(M)$, we focus on the 
	Amelino-Camelia dispersion relation  with $n = 2$ and $\eta > 0$.
	 Thus we get
	\begin{equation}
	ImI_{+}-ImI_{-}=\int\frac{4}{\sqrt{1-\eta x^2}\Delta}\sqrt{\omega^2r^4+2a\omega^2 r^2-aM\omega j\sqrt{1-\eta x^2}r+C_{1}^2-\Delta C_{2}(r)}dr
	.
	\end{equation}
	Hence the Hawking temperature is expressed by mass $M$
	\begin{equation}
	\widetilde{T}(\eta)=T_{h} \sqrt{1-\frac{\eta  \left(m^2 \left(\sqrt{M^2-a^2}+M\right)^2+1\right)}{\left(\sqrt{M^2-a^2}+M\right)^2+\eta }}
	,
	\end{equation}
	or in the form
	\begin{equation}
	\widetilde{T}(r_{+})=T_{h} \sqrt{1-\frac{\eta  (m^2 r_{+}^2+1)}{r_{+}^2+\eta }},
	\end{equation}
	 which happens to be 
	\begin{equation}
	\widetilde{T}=T_{h}\sqrt{1-\eta x^2}=T_{h}g(x).
	\end{equation}
	 In fact, Hawking temperature corrections of many other rainbow black holes have the same expression as above\cite{Morais:2021xmw,Feng:2021zsq}. Subsequently,  using the first law of black hole thermodynamics
	 \begin{equation}\label{10}
	 TdS=dM+\Omega _{h}dJ,
	 \end{equation}
	 we find that the entropy of the black hole is
	 \begin{equation}\label{10}
	 S=\int \frac{a+1}{T} dM,
	 \end{equation}
	 To numerically investigate the Hawking temperature $T (M)$ and the entropy $S (M)$ in ST, we 
	 plot $T (M)$ and $S (M)$ for various values of $\eta$ in Fig. (2), where we focus on the Amelino-Camelia dispersion relation with $n = 2$ and $\eta > 0$. In the following discussions, we take $m = 0.01, a=0.1$ to discuss the thermodynamic properties of Kerr black hole. 
	 
	 Varying the effect parameter $\eta$, the left panel of Fig.(2) shows that 
	 the black hole temperature decreases with increasing $\eta$, which implies that the rainbow effects would slow down the evaporation of the black hole in ST frame. On the other hand, the right panel of (2)  shows that the black hole entropy increases with increasing $\eta$ in the ST rainbow Kerr case, which means the black hole tends to store more information.
		\begin{figure}[t]\label{fig2}
		\centering
		\subfigure[Hawking temperature of different $\eta$.]{
			\includegraphics[scale=0.6]{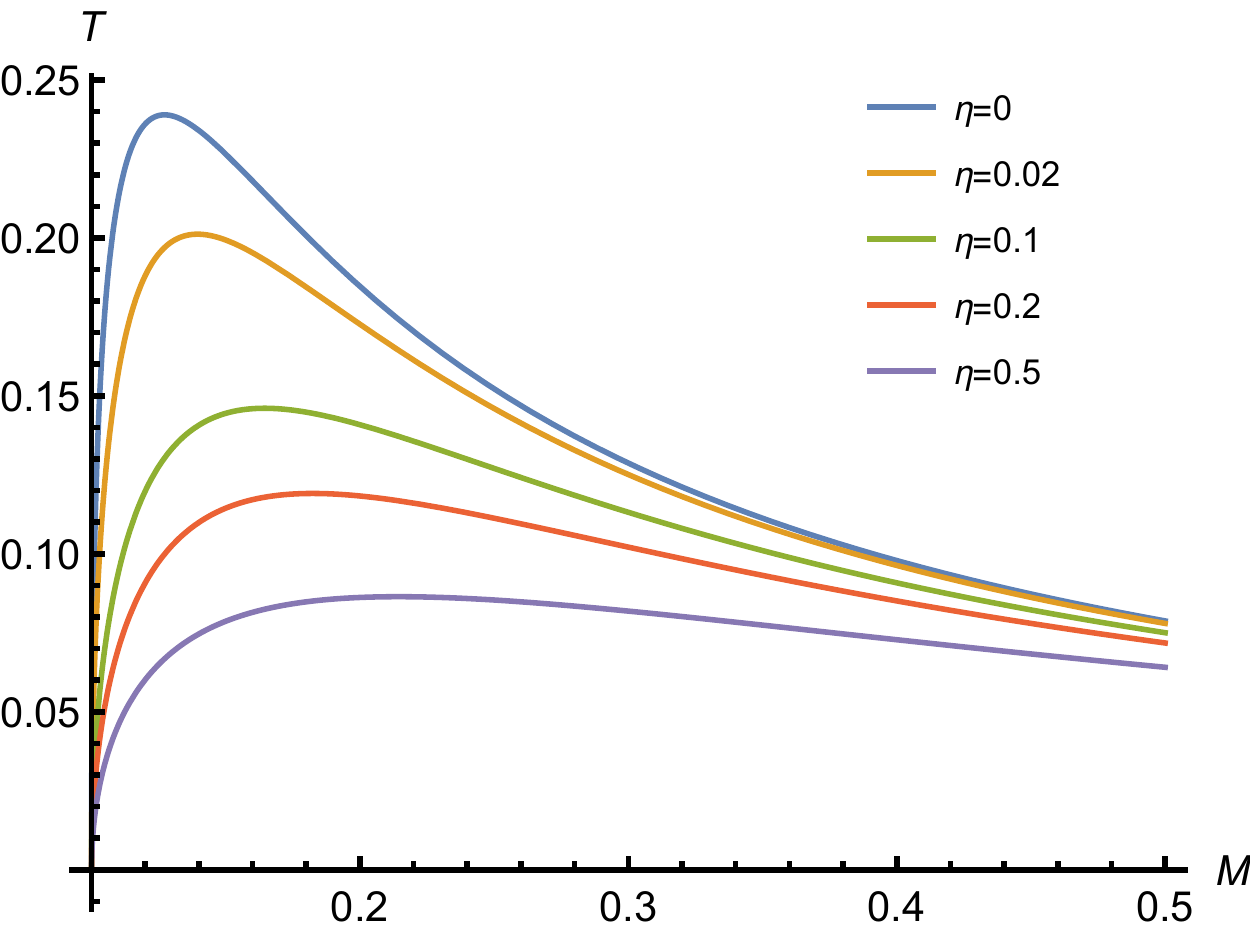}}
		\quad
		\subfigure[Entropy of different $\eta$.]{
			\includegraphics[scale=0.6]{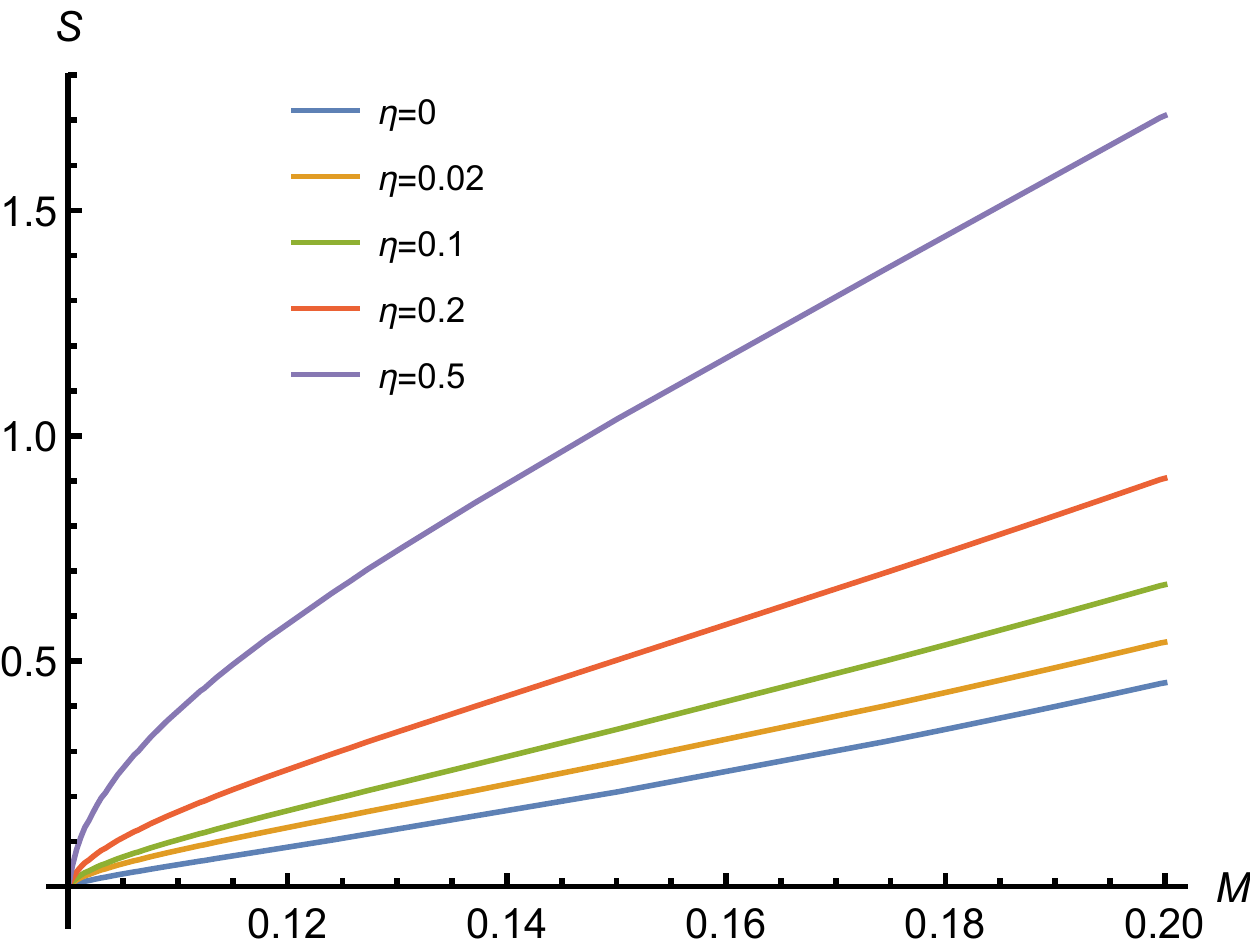}}
		\caption{Hawking temperature $T(M)$ and entropy $S(M)$ of a ST rainbow Kerr black hole with various $\eta$ for $n=2$.}
	\end{figure}
  \begin{figure}[t]\label{fig3}
  		\centering
  		\subfigure[Hawking temperature of different $n$ with $\eta=0.9$.]{
  			\includegraphics[scale=0.6]{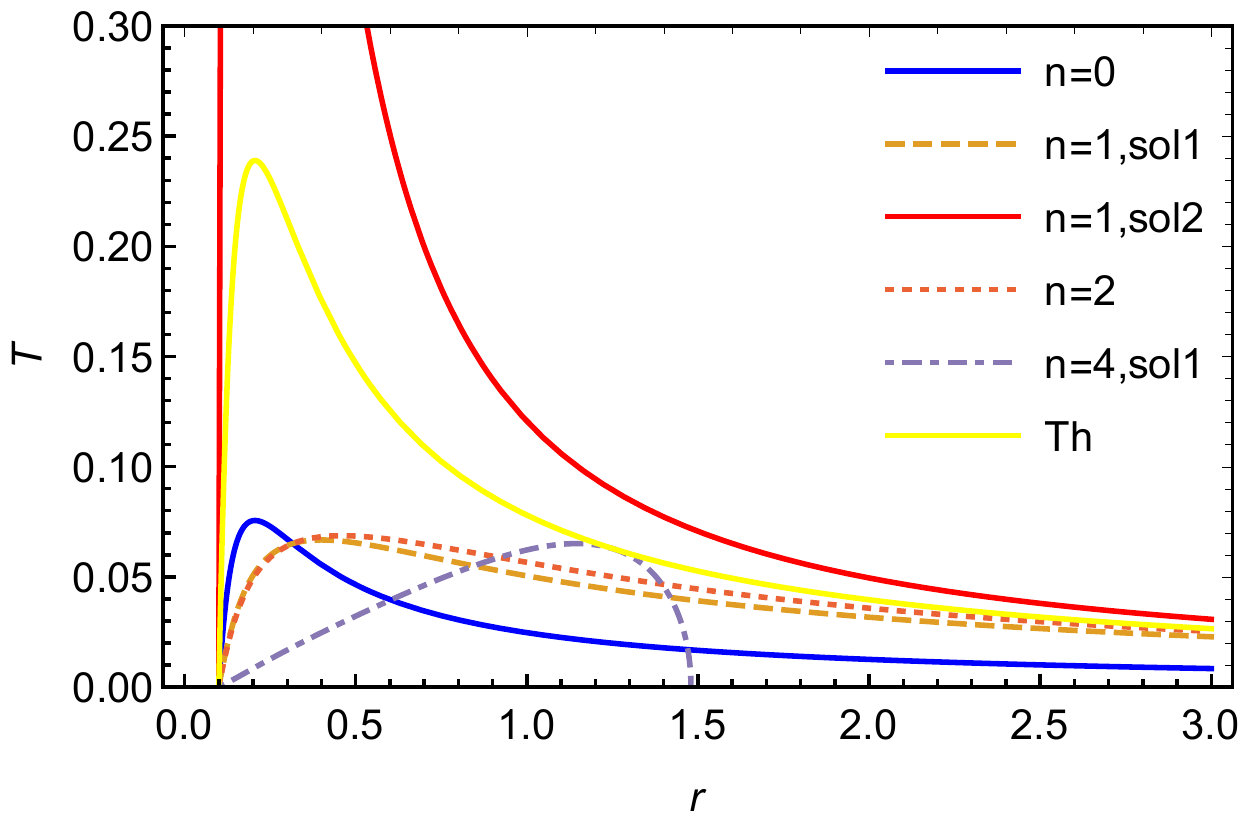}}
  		\quad
  		\subfigure[Hawking temperature of different $n$ with $\eta=-0.9$.]{
  			\includegraphics[scale=0.6]{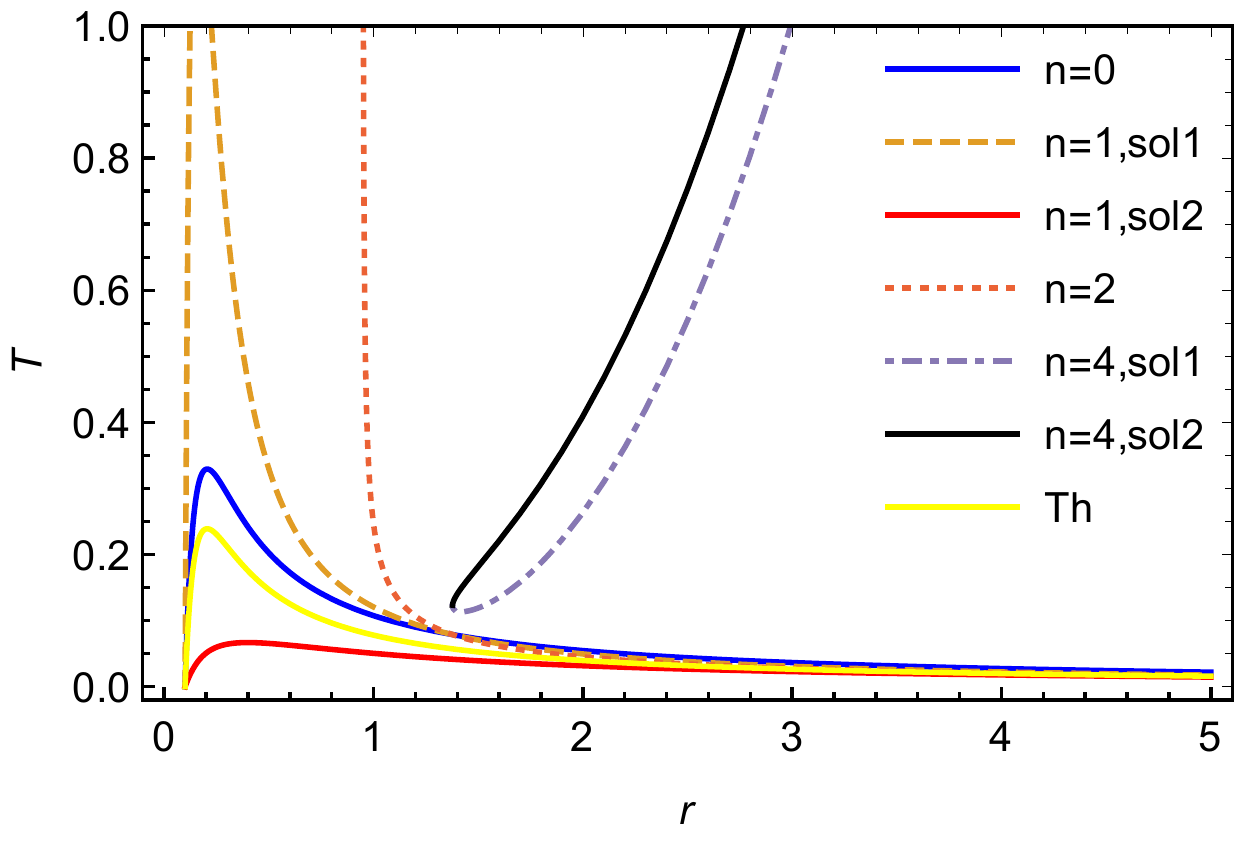}}
  		\caption{Hawking temperature $T(r_{+})$ of a ST rainbow Kerr black hole with various $n$ for $n=\pm0.9$.}
  \end{figure}
  
  Varying the exponential parameter $n$, we set $\eta=\pm0.9$ to discuss. To simplify the discussion, we visualize Hawking temperature and entropy in the ST frame in Fig.(3), where $T_{h}$ means the standard Hawking temperature.
  
  For $\eta>0$, the temperatures have concave downward trends for general relativity and also for $n = 0$. The shape of the curve is different for different exponents, but
  their behavior is similar except occassion ($n=4, sol1$), expressed by blue dotted line. Through the figure, we find the size of black holes is limited in a definite range for $(n=4, sol1$), while occassion ($n=4, sol2$) does not exist.
  
  For $\eta<0$, all studied cases until $n < 4$ leads to an zero temperature as the radius
  approaches infinity. For $n = 4$ we have two solutions with a
  bound to the horizon radius. Two curves meet and terminate at the same minimum radius of the event horizon, and the temperatures increase with increased $r$. This is a confusing new behavior,
  as the diagram point out that the black hole at the same temperature have two defferent solutions of $r$ that meet the physical requirements, which implies a fixed temperature does not guarentee a determinate event horizon.
  
  To further explore the stability and phase transition of black holes, we derive the specific heat which is defined by equation
  \begin{align}\label{18}
  C_{k}=\frac{dM}{dT}=\frac{dM}{dr}(\frac{dT}{dr})^{-1},
  \end{align}
  and disscuss its properties of defferent solutions in detail.
   Considering the complex analytic expression, the specific heat is visualized and depicted in Fig.(4).

  For $\eta>0$, Curves' behavior are similar to the original case $n=0$, except occassion ($n=4, sol1$), expressed by purple dotted line. For solution ($n=4, sol1$), the specific heat
  only exists at a certain distance, and the asymptotic behavior occurs at a larger $r$,  which is consistent with the temperature image.

  For $\eta<0$, new cases occur for $n = 4$ and $n = 2$. For ($n=4, sol1$), the trend of curve is oppsite to $\eta>0$ case. For $n=4, sol2$, it is always positive and terminates at the same starting point with  ($n=4, sol1$). For $n=2$, it is always negative.
 
  With these results, we can study the thermodynamic stability, phase transitions, and critical points. As the specific heat of a black hole is positive, it's thermodynamically stable. Black holes with negative specific heat, on the other hand, are unstable. In this case, first or second-order phase transitions are possible. A first-order phase transition occurs when specific heat vanishs, and the divergent point of specific heat indicates the second order phase transition.
    \begin{figure}[t]\label{fig4}
  	\centering
  	\subfigure[the specific heat with $\eta=0.9$.]{
  		\includegraphics[scale=0.6]{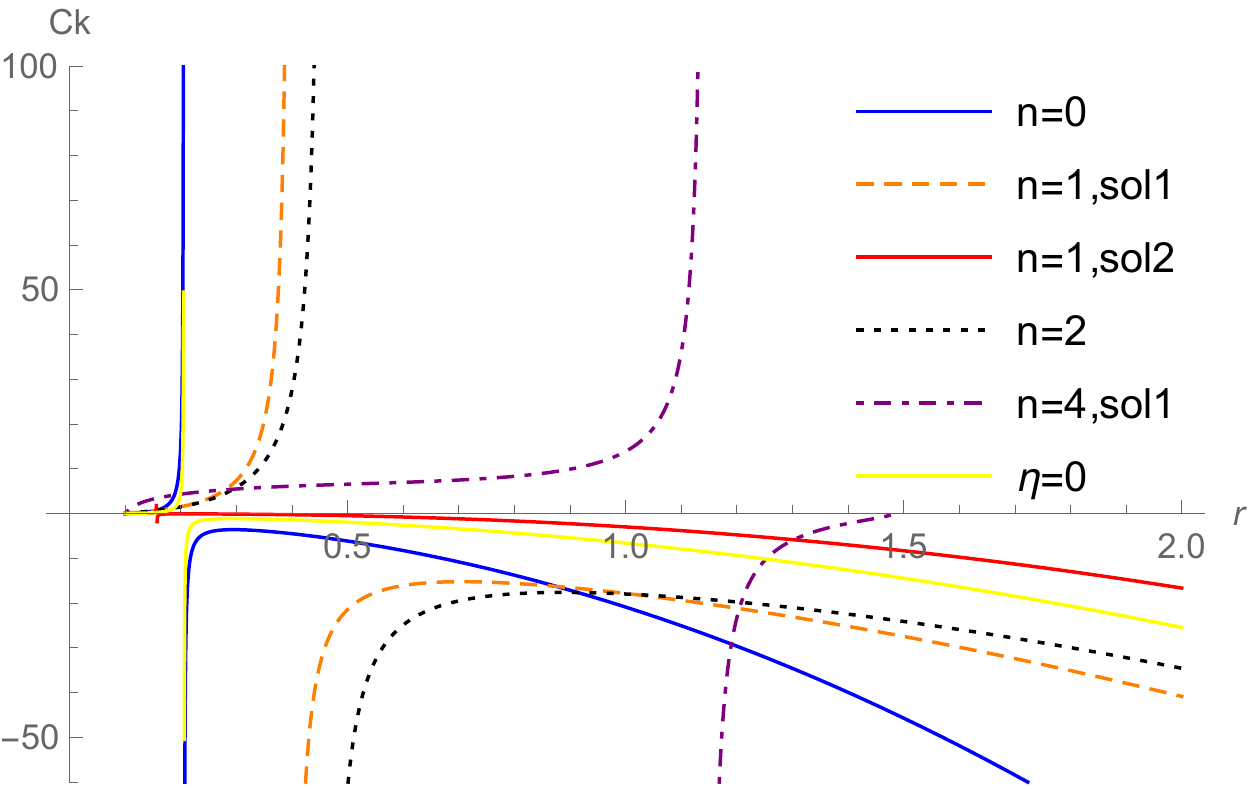}}
  	\quad
  	\subfigure[the specific heat with $\eta=-0.9$.]{
  		\includegraphics[scale=0.6]{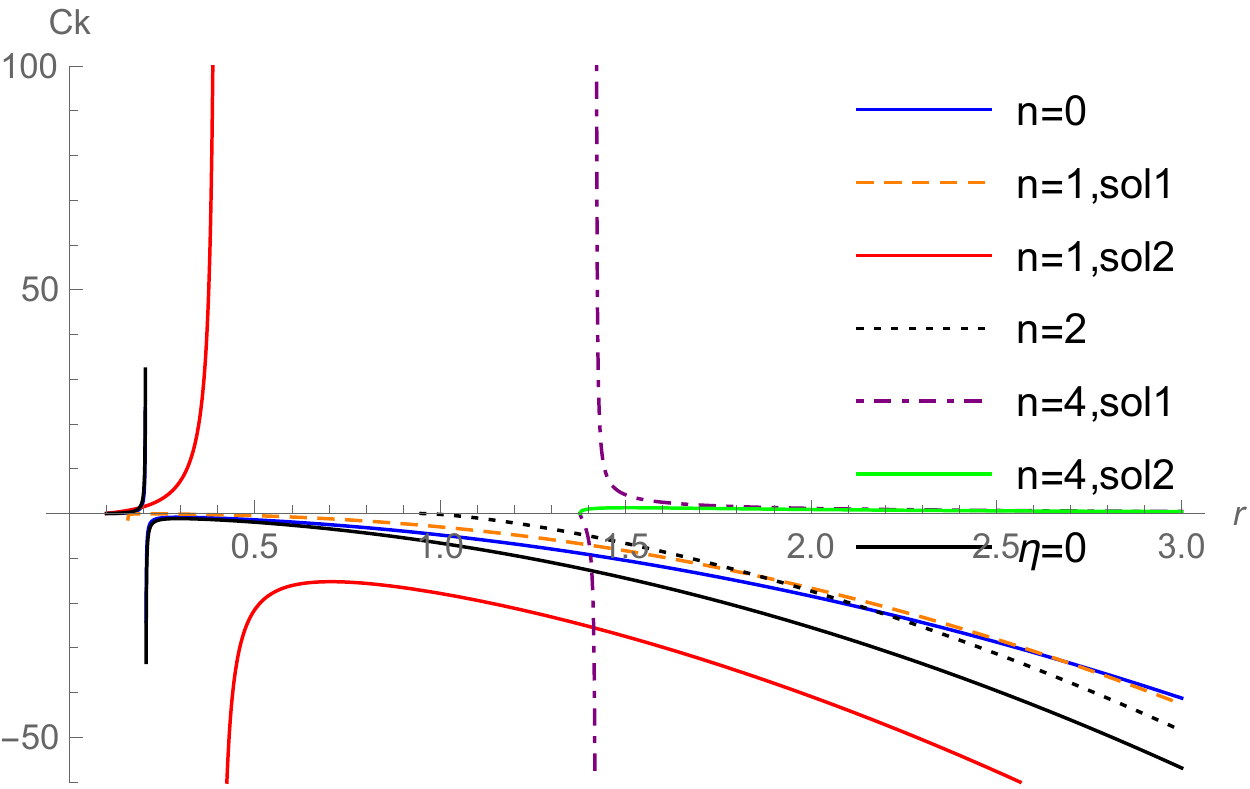}}
  	\caption{Plots of the modified Kerr black hole's specific heat $C_{k}(r)$ by horizon radius $r$ with various values of $n$ with rainbow gravity.}
  \end{figure}
  \section{Thermodynamics in free-fall frame}
     \subsection{FF frame rainbow Kerr black hole}  
     To study the thermodynammics of Kerr black holes in FF frame, we start with four dimensional velocity
    \begin{align}
    u^{t}&=-\frac{g_{\phi\phi} E+g_{t\phi}L}{(g_{t\phi}^2-g_{tt}g_{\phi\phi})m},\\
    u^{\phi}&=-\frac{-g_{b\phi} E-g_{tt}L}{(g_{t\phi}^2-g_{tt}g_{\phi\phi})m},\\
    u^{\theta}&=g_{\theta\theta}u^{\theta}.
    \end{align}
    and geodesic equations, along with considering the $ p_{r} $ and $p_{\phi}$ is conserved along the geodestic\cite{Carter:1968rr}
    \begin{align}
    g^{tt}u_{t}u_{t}+2g^{t\phi}u_{t}u_{\phi}+g^{\phi\phi}u_{\phi}u_{\phi}+g^{rr}u_{r}u_{r}+g^{\theta\theta}u_{\theta}u_{\theta}=-1,\\
    (u^{r})^2=-g^{rr}(g^{tt}u_{t}u_{t}+2g^{t\phi}u_{t}u_{\phi}+g^{\phi\phi}u_{\phi}u_{\phi}+g^{\theta\theta}u_{\theta}u_{\theta}).
    \end{align}
    Therefore we get the 4-dimention velosity of the particle
    \begin{align}
    u^{\mu}=(-\frac{E}{m},\quad \sqrt{\frac{\rho ^2 \left(\frac{\Theta ^2}{\rho ^2}+\frac{E^2 \Sigma ^2}{\Delta  m^2 \rho ^2}-1\right)}{\Delta }},\quad \frac{L}{m},\quad \Theta (\theta)).
    \end{align}
    The energy-dependent rainbow counterpart for the energy-independent
    metric can be obtained by equivalence principle ,
    which gives
    \begin{align}\label{0}
     d\widetilde{s}^2=\widetilde{g}_{\mu\nu}dx^{\mu}\otimes dx^{\nu}=
     (\frac{1}{g^2(\frac{E}{Ep})}-\frac{1}{f^2(\frac{E}{Ep})})e_{0}\otimes e_{0}+\frac{ds^2}{g^2(\frac{E}{Ep})},
    \end{align}
    where the energy-dependent orthonormal frame fields are
    \begin{align}
   \widetilde{e}_{0}=\frac{e_{0}}{f(\frac{E}{Ep})},\quad\widetilde{e}_{i}=\frac{e_{i}}{g(\frac{E}{Ep})}.
    \end{align}
    The time component of the orthonormal frame anchored
    to the radiated particle is then given by
    \begin{align}
    e_{0}=u_{\mu}dx^{\mu}=-\frac{E}{m}dt+\sqrt{\frac{\rho ^2 \left(\frac{\Theta ^2}{\rho ^2}+\frac{E^2 \Sigma ^2}{\Delta  m^2 \rho ^2}-1\right)}{\Delta }}dr.
    \end{align}
   
   To solve Hamiton-Jacobi equation, we need find $\widetilde{g}^{\mu\nu}$, which is the inverse of $\widetilde{g}_{\mu\nu}$. We solve the equation to get the momentum  $Pr$ 
   \begin{align}
   -2g^{t\phi}\omega j+g^{\theta\theta}\Theta^2+g^{\phi\phi}j^2+g^{tt}\omega^2+m^2F-2g^{rr}   
   P_{r}^2+2g^{r\phi}P_{r} j=0.
   \end{align}
   Two solution about $Pr$ is obtained which is too long to put here, named $Pr_{+}$ and $Pr_{-}$. They could be expressed as follows
   \begin{align}
   Pr_{+}-Pr_{-}=\frac{2 \sqrt{A B \rho ^2}}{\Delta  F g \rho ^2},
   \end{align}
   and we define
   \begin{align}
   A&=a^2 \sin ^2\theta (4 M^2 r^2-\Sigma ^2)((\Delta  \Theta ^2+E^2 \Sigma ^2) (f^2-g^2)+\Delta  g^2 m^2 \rho ^2)\nonumber\\
   &+\Delta  \Sigma ^2 (f^2 (\Delta  \Theta ^2+E^2(\Sigma ^2-\rho ^4))+g^2 (E^2 (\rho ^4-\Sigma ^2)+\Delta  (m^2 \rho ^2-\Theta ^2))),\\
   B&=a^2 f^2 m^2 \sin ^2\theta  (\Sigma ^2-4 M^2 r^2) (g^2 \Theta ^2+m^2 \rho ^2)-\Sigma ^2(\Delta  f^2 g^2 \Theta ^2 m^2)\nonumber\\
   &-\Sigma ^2 (\rho ^2(\Delta  f^2 m^4+E^2 g^2 \Theta ^2 (g-f) (f+g))+E^2 m^2 \rho ^4 (-((f^2-1) g^2+f^2))),\\
   F&=a^2 f^2 m^2 \sin ^2\theta  \left(4 M^2 r^2-\Sigma ^2\right)+\Sigma ^2 \left(E^2 \rho ^2 \left(g^2-f^2\right)+\Delta  f^2 m^2\right).
   \end{align}
   Then we can study the thermodynamics in FF frame.
\subsection{Thermodynamics in FF rainbow Kerr black holes}   
    In FF frame, let's set $n$ equal to $2$ for convinience, and the metrics becomes
    \begin{align}
    ds^2=& (-\frac{\Delta -a^2 \sin ^2\theta }{\rho ^2 (1-\eta  x^2)}+(\frac{E}{m})^2 (\frac{1}{1-\eta  x^2}-1))dt^2-\frac{E (\frac{1}{1-\eta  x^2}-1)}{m} u_{r} drdt\nonumber\\
    &+(u_{r}^2 (\frac{1}{1-\eta  x^2}-1)+\frac{\rho ^2}{\Delta  (1-\eta  x^2)})dr^2-\frac{2 a M r }{\rho ^2(1-\eta  x^2)}\sin ^2\theta dtd\phi\nonumber\\
    &+\frac{\Sigma ^2}{\rho ^2(1-\eta  x^2)} \sin ^2\theta d\phi^2+\frac{\rho ^2}{1-\eta  x^2}d\theta^2.
    \end{align}
    \begin{figure}
    	\centering
    	\subfigure[Hawking temperature of different $\eta$.]{
    		\includegraphics[scale=0.6]{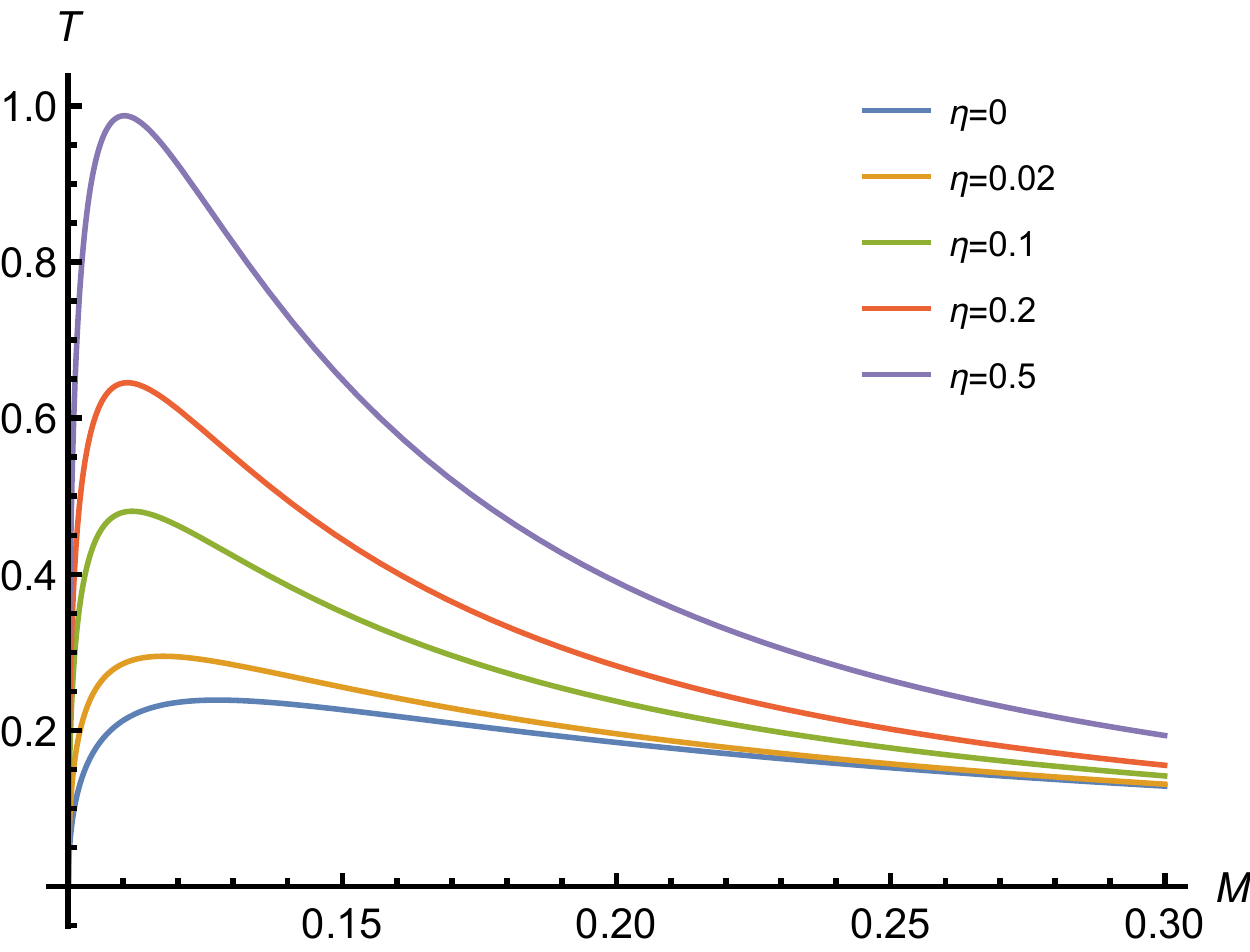}}
    	\quad
    	\subfigure[Entropy of different $\eta$.]{
    		\includegraphics[scale=0.6]{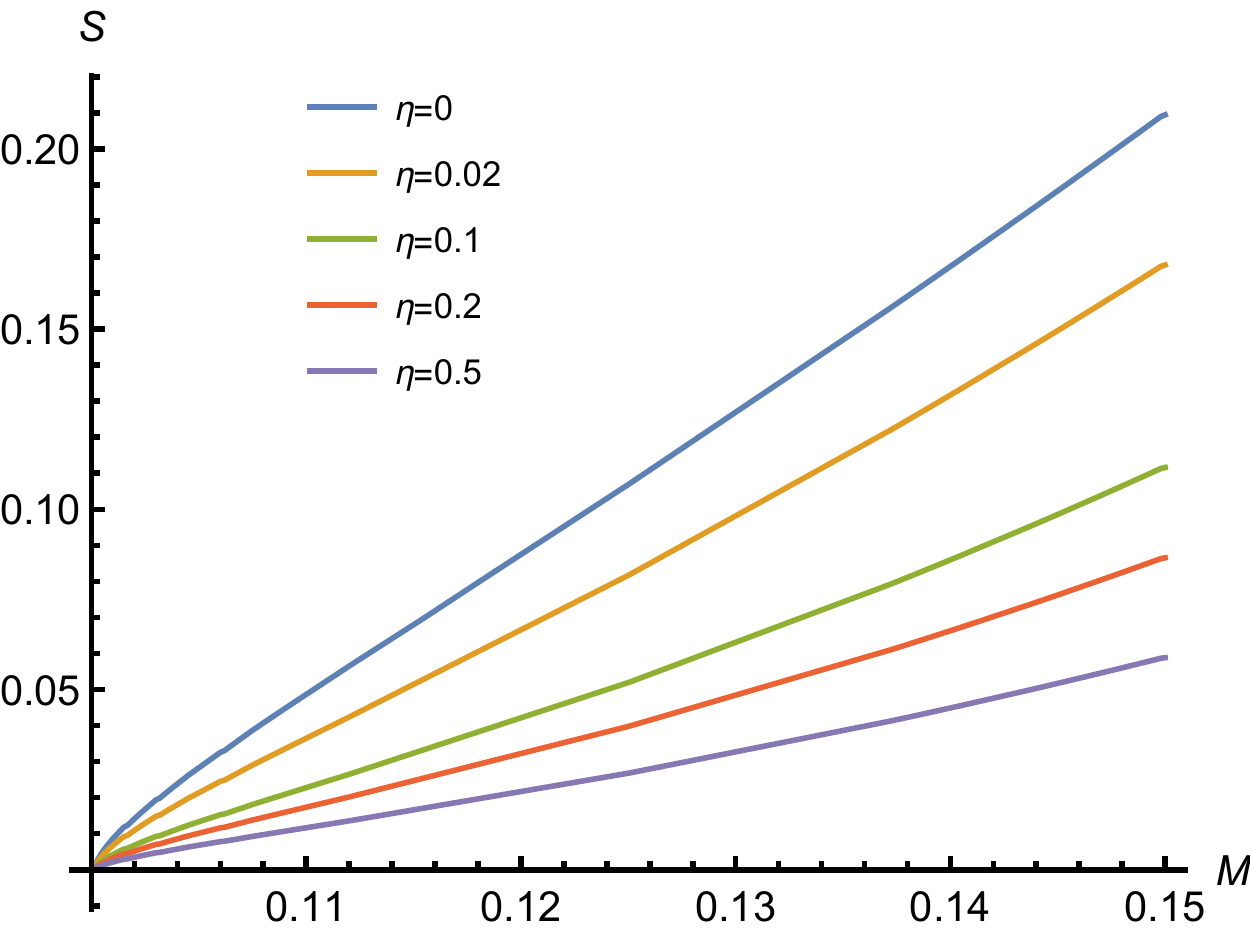}}	
    	\caption{Plots of the Hawking temperature $T(M)$ and entropy $S(M)$ of a FF rainbow Kerr black hole various  $\eta$ for $n=2$.}
    \end{figure}
    According to (\ref{16})
   	we apply the residue theorem of semi-circle. Since the terms of $\theta$ are multiplied by the higher order terms of $\Delta$, the numerator terms of $\Theta$ are eliminated. Take equation(\ref{2}) and equation(\ref{9}) into account, the Hawking temperature expressed by mass $M$ is obtained through the discussion similar to previous one. To simplify, we express it as a function of $r_{+}$ and $r_{-}$
   	\begin{align}
   	\widetilde{T}(\eta)=T_{h}\frac{(1-\frac{\eta(m^2r_{+}^2+1)^2(r_{+}^2+r_{-}r_{+})^2}{2(\eta +r_{+}^2)^2})}{\sqrt{1-\frac{\eta(m^2 r_{+}^2+1)}{\eta +r_{+}^2}}},
   	\end{align}
   	which is equal to
   	\begin{align}
   	\widetilde{T}(\eta)=T_{h}\frac{1-\frac{(r_{+}^2+r_{-}r_{+})^2}{2}(1-g^2)^2}{g},
   	\end{align}
   	
   	Through the similar method of section \uppercase\expandafter{\romannumeral4}, we obtain the modified Hawking temperature and entropy in FF frame by analytical and numerical method, as shown in Fig.(5) with $n = 2$ and $\eta > 0$.
   	The left panel of Fig.(5) shows that 
   	the black hole temperature increases with increasing $\eta$, which implies that the rainbow effects in FF frame would speed up the evaporation of the black hole. On the other hand, the right panel of Fig.(5) shows 
   	the black hole entropy decreases with increasing $\eta$. Therefore, the 
   	black hole tends to store less information when the rainbow effects are turned on.
   	 \begin{figure}
   		\centering
   		\subfigure[Hawking temperature of  $\eta=0.1$.]{
   			\includegraphics[scale=0.6]{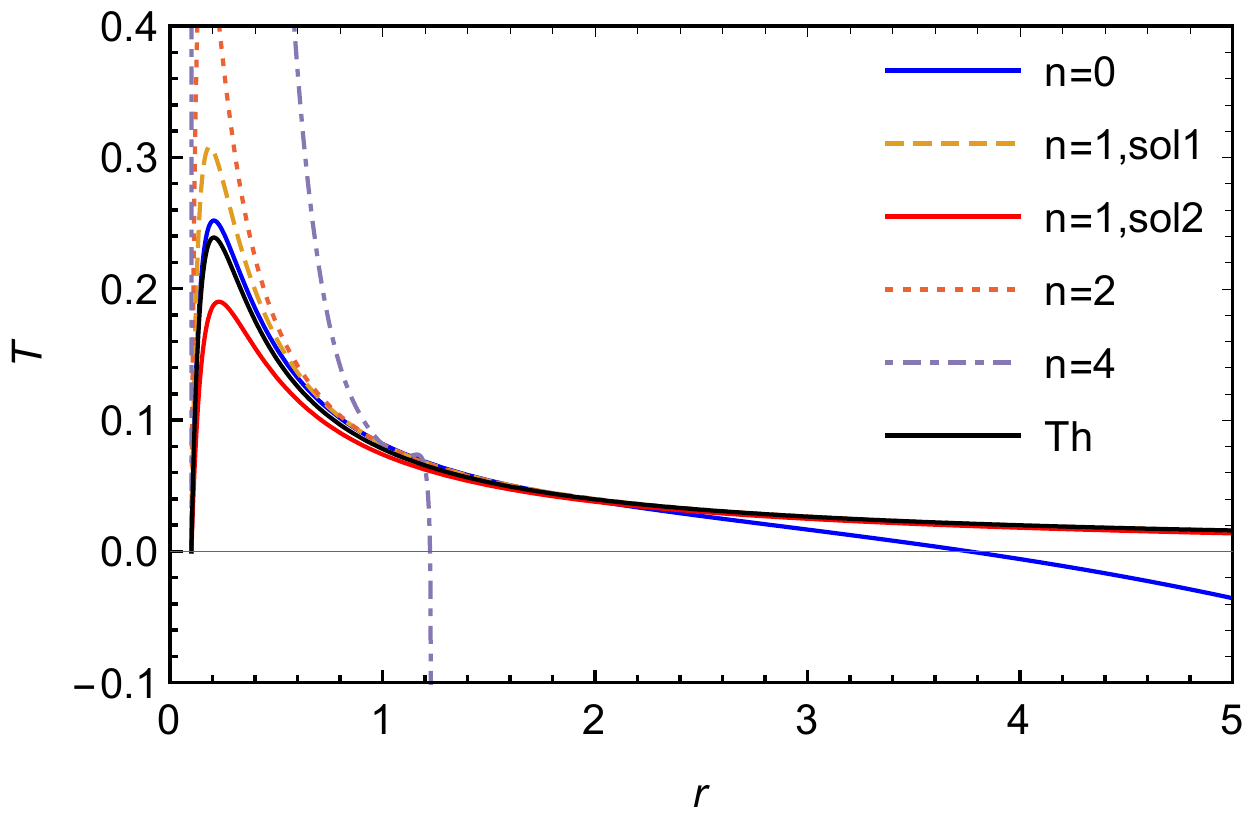}}
   		\quad
   		\subfigure[Hawking temperature of $\eta=-0.1$.]{
   			\includegraphics[scale=0.6]{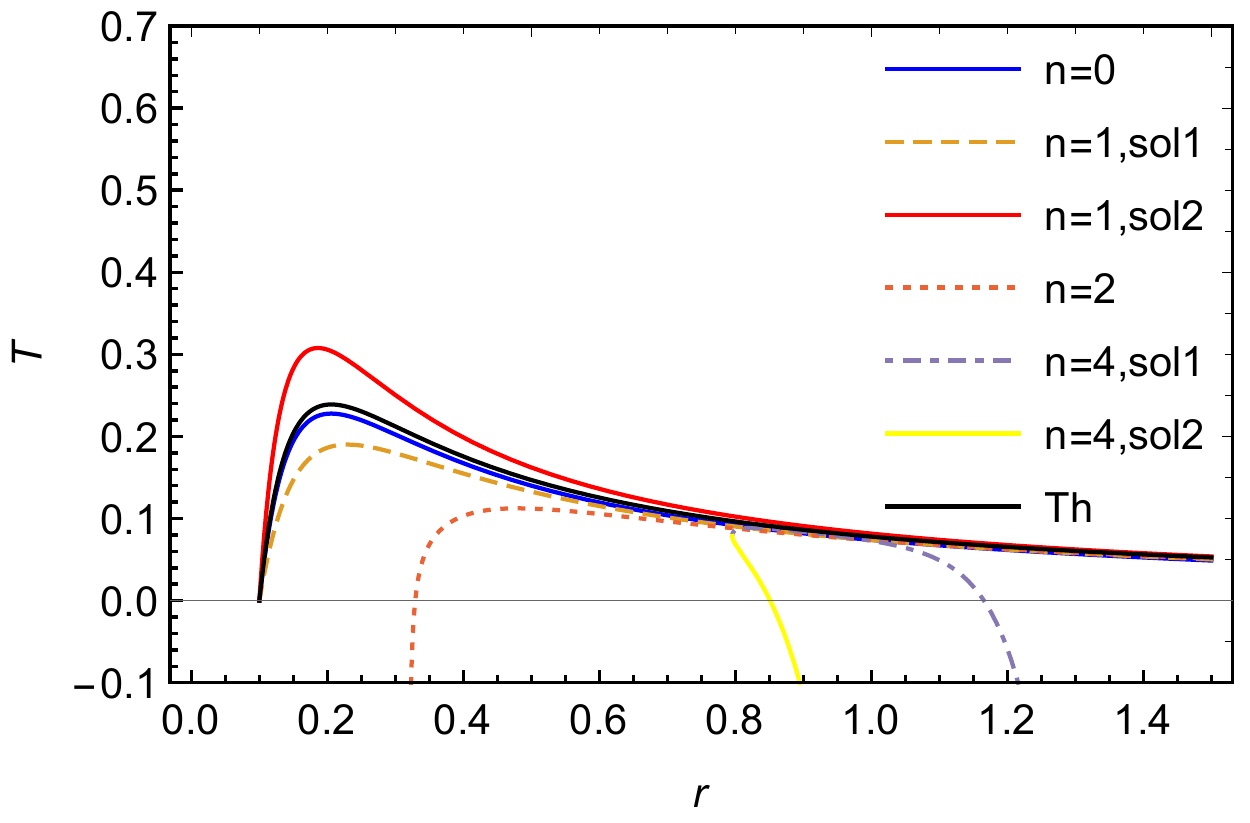}}	
   		\caption{The Hawking temperature $T(r)$ of a FF rainbow Kerr black hole with various $n$.  $T_{h}$ represents the original Hawking temperature.}
   	\end{figure}
 
    We also plot the modified Hawking temperature and entropy for defferent $n$ in FF frame in Fig.(6) and Fig.(7), where the parameters of rainbow gravity are more widely chosed.
    For $\eta =0.1$, all studied cases except $n = 4$ and $n = 2$ leads to an zero temperature as the radius
    approaches infinity. Specificly, the temperatures of $n = 4$ and $n = 0$ turn negative with increasing $r$ .
    For $\eta=-0.1$ all studied cases until $n = 4$ have a concave downward trends. However, for $n = 2$, the temperature begins from negative infinity. As for two solutions of $n = 4$, they begin at the same point and tend to negative infinity.
    In both case, the abnormal negative temperatures occur, which means $n=4$ is not physically accepted. Morover, $n=2$ case should be considered only with a positive $\eta$. 
   
    In FF frame, the entropy of black hole is also modified differently due to the choice of $n$, and we use numerical method to draw a graph of entropy with respect to mass $r$ in Fig.(7).
    
    \begin{figure}[t]
    	\centering
    	\subfigure[Entropy of  $\eta=0.1$.]{
    		\includegraphics[scale=0.6]{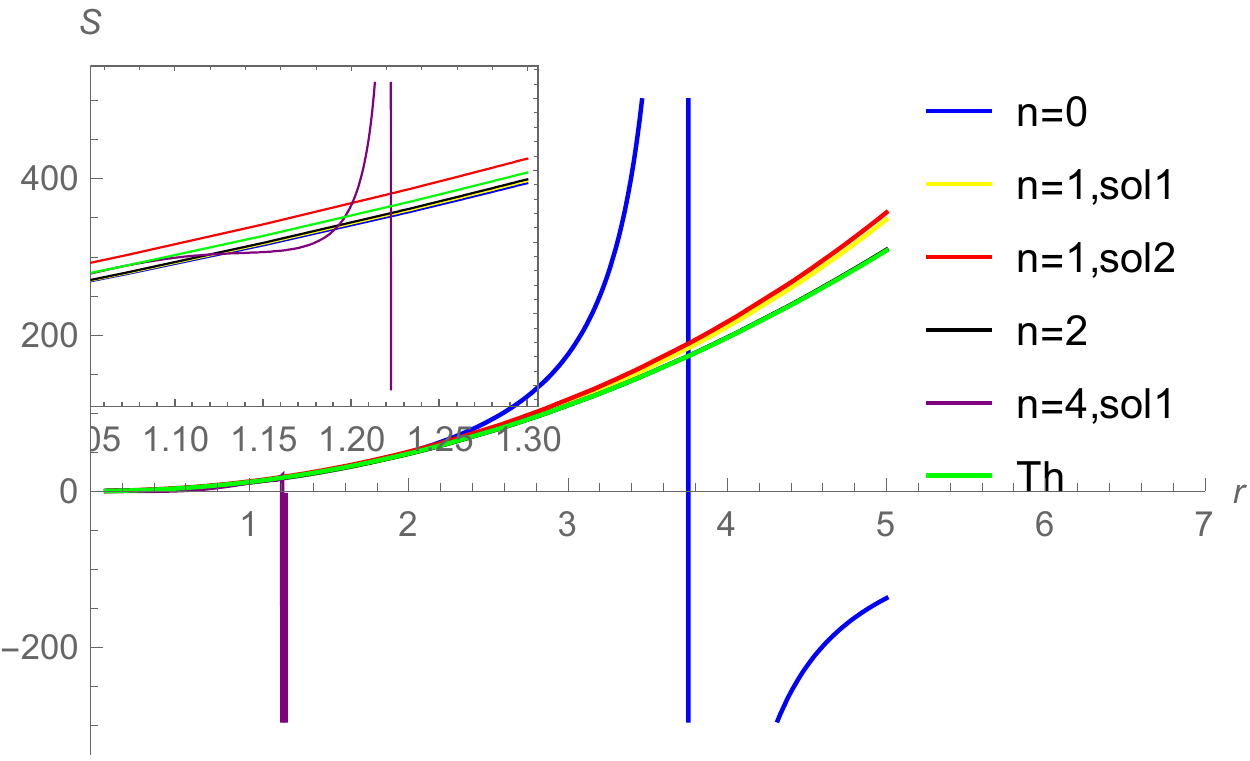}}
    	\quad
    	\subfigure[Entropy of $\eta=-0.1$.]{
    		\includegraphics[scale=0.6]{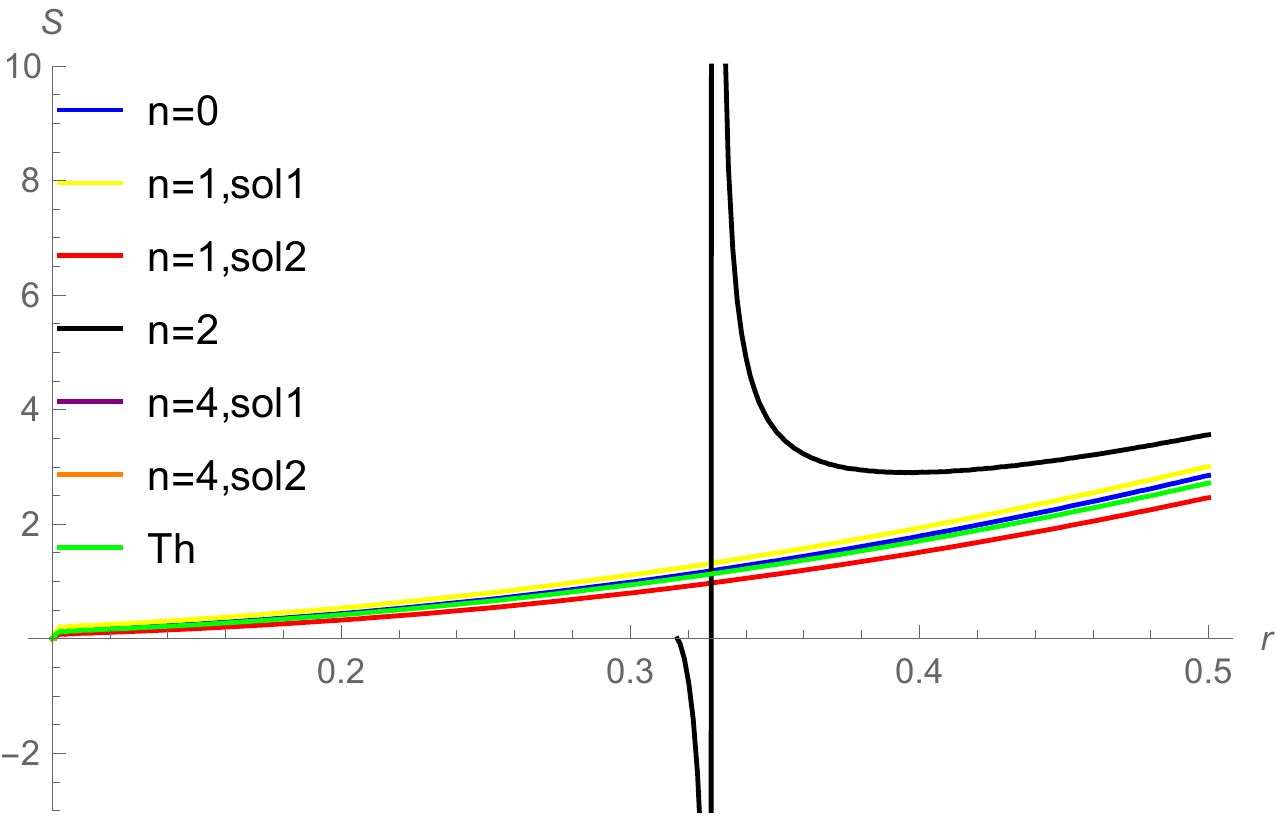}}	
    	\caption{Plots of the entropy $S(r_{+})$ of a FF rainbow Kerr black hole with various $n$.}
    \end{figure}
    When $\eta =0.1$, all solutions except $n=0$ and $n=4$ have a normal growth trend similar to that in ST Frame. As the temperature curves of $n=0$ and $n=4$ have a positive to negative trend, asymptotic behavior occurs at critical points, which is not feasible physically.
    When $\eta =-0.1$, as the temperature of $n=2$ and $n=4$ all have a positive to negative transition, they also have asymptotic behaviors. Similar to the previous discussion, $n=2$ and $n=4$ are invalid for negative $\eta$.
  
    Through formula (\ref{18}), we can caculate the specific heat of Kerr black hole in FF frame. As operated in the previous section, we visualize the specific heat to discuss its properties rather than using lengthy formulas. 
    
    For $\eta=0.1$, all solutions except $n=4$ are different from original case numerically, but their behavior is similar.
    In addition, in the limit of a small $r$, the specific heat of $n=0$ and $\eta=4$ is almost the same and the curves in the left panel of Fig.(8)  overlap. For $n=4$ case, two asymptotic behaviors occur and shift in the positive direction of the $r$ axis. Morover, it suggests that the size of black holes is limited, as the specific heat only exists in a limited range.
    For $\eta=-0.1$, the solutions of $n=2$ and $n=4$ are found nontrival again, compared with the discussion in section \uppercase\expandafter{\romannumeral2}.
    \begin{figure}
    	\centering
    	\subfigure[Specific heat in FF frame of  $\eta=0.1$.]{
    		\includegraphics[scale=0.6]{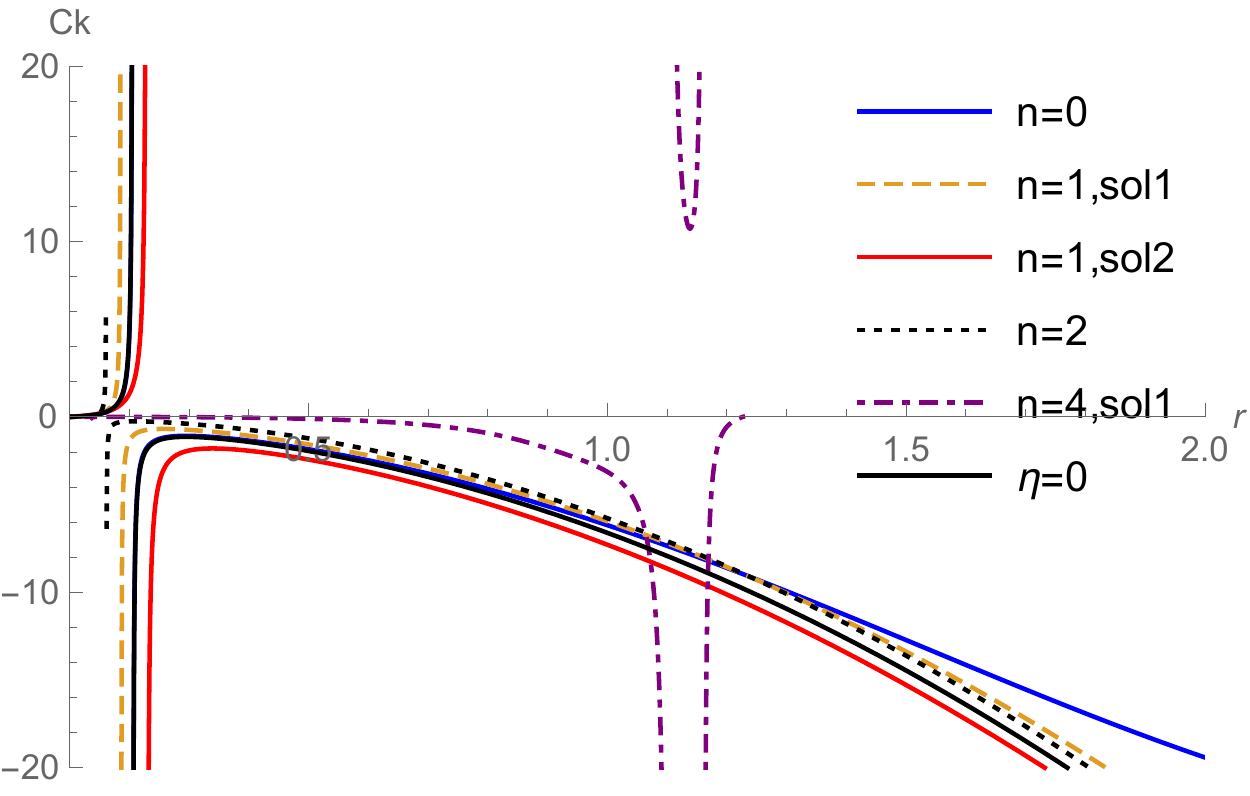}}
    	\quad
    	\subfigure[Specific heat in FF frame of $\eta=-0.1$.]{
    		\includegraphics[scale=0.6]{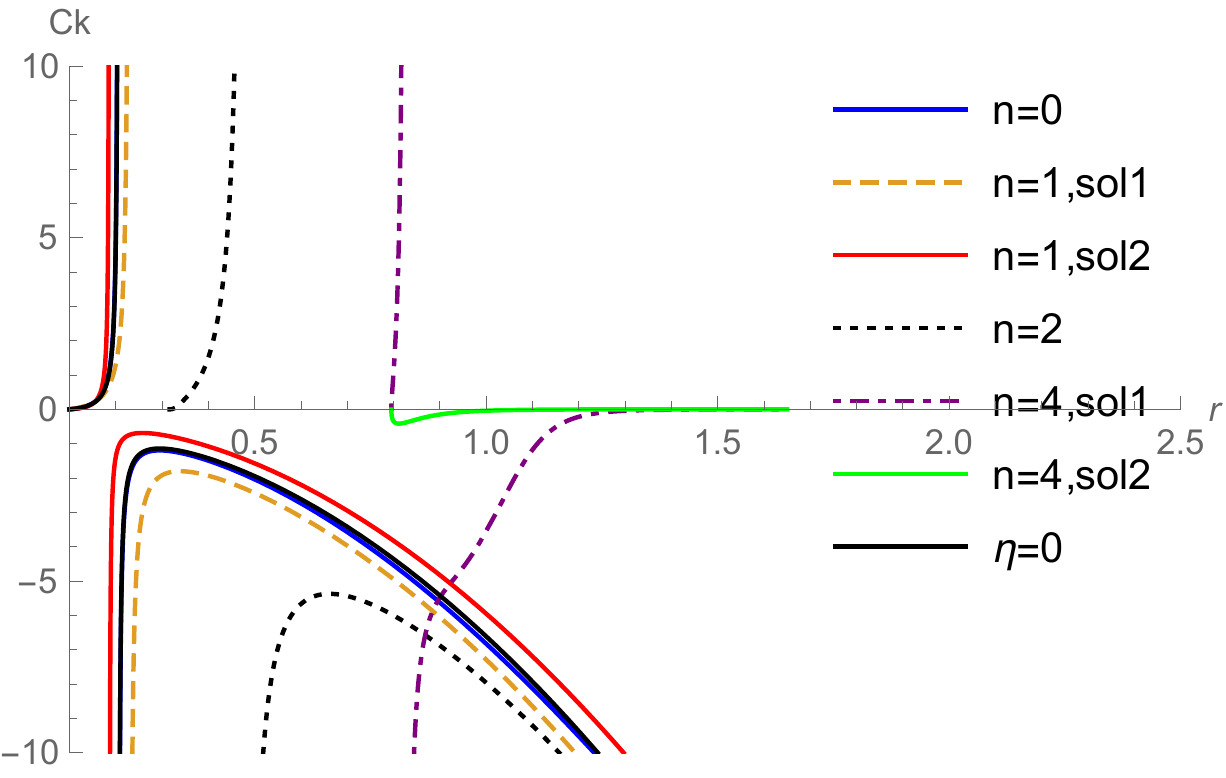}}	
    	\caption{The specific heat $C_{k}(r)$ of a FF rainbow Kerr black hole with various $n$ .}
    \end{figure}
  \section{Discussion and conclution}

    In this paper, we consider rainbow Kerr black holes in ST frame and FF frame, and the metric of the FF rainbow Kerr black hole was first derived. Then, we use the Hamilton-Jacobi method to obtain the effective Hawking temperature of rainbow Kerr black holes in ST frame and FF frame, which are shown to depend on the energy and angular momentum of radiated particles. However, the effective Hawking temperature of a ST rainbow Kerr black hole is always finite and positive, while the effective Hawking temperature of a FF rainbow Kerr black hole may be negative depends on the choice of $n$. Focusing on the Amelino-Camelia 
    dispersion relation, we numerically study the temperature and 
    entropy of the ST and FF rainbow Kerr black holes. The corresponding results and implications are summarized in Table 2.
    \begin{table}[t]\label{50}
    	\centering
    	\caption{Results and implications for the Hawking temperature and the black hole entropy for the FF and ST rainbow Kerr black holes in the subluminal case ($n=2,\eta>0$).}
    	
    	\begin{tabular}{p{2.5cm}|p{7cm}|p{7cm}}
    		\hline
    		&FF rainbow Kerr black hole&ST rainbow Kerr black hole   \\
    		\hline
    		 Temperature&The rainbow effects increase the temperature, which leads to a more 
    		 violent death with a nonzero terminal temperature.&The rainbow effects decrease the temperature, which leads to a more 
    		 peaceful death with a zero terminal temperature.     \\
    		\hline
    		 Entropy& The rainbow effects decrease the entropy, which means the black 
    		 hole can store less information.&  The rainbow effects increase the entropy, which means the black 
    		 hole can store more information.    \\
    		\hline
    	\end{tabular}
    \end{table}
    
   To investigate the effects of parameter $n$, the physical feasibility of different solutions and the relative modification of feasible solutions are summarized and discussed. We find that some $n$ choices in the FF frame may lead to negative Hawking temperature and the discontinuous changes in entropy, but not in the ST frame. Discussion of thermodynamic properties in the FF frame can help us rule out some non-physical cases. For example, when $n=2$, $\eta$ can only be positive.
  
   The Schwarzschild black holes based on rainbow gravity has been discussed in the ST and FF cases\cite{Mu:2015qna,Tao:2016baz}. For the ST and the FF rainbow Schwarzschild black holes, the rainbow gravity tends to decrease the Hawking temperature and increase the entropy in the subluminal case. However, our results show that the Hawking temperature and the entropy of the ST and the FF rainbow Kerr black hole behave rather differently. It seems that the effects that the rainbow gravity has are quite frame-dependent. So it would be interesting to study the thermodynamic properties of various black holes of different theories of gravity in different rainbow models, which may help us explore the effects of quantum gravity. In addition, the boundary conditions are critical to the metric of a black hole. It would be interesting to study the effects of FF rainbow Kerr black holes in cavity in the future, which may help us explore more about the rainbow gravity.
    
    \begin{acknowledgements}
    	We are grateful to Haitang Yang and Peng Wang for useful discussions. This work is supported
    	in part by NSFC(Grant No.11375121,11747171,11747302 and11847305). Natural Science Foundation of Chengdu University of TCM(Grants No.ZRYY1729 and ZRON1656). Discipline Talent Promotion Program of Xinglin Scholars (Grant No. ONXZ2018050) and The Key Fund Project for Education Department of Sichuan Province(Grant No.18ZA0173). Sichuan University Students Platform for Innovation and Entrepreneurship Training Program(Grant No.C2019104639). The authors contributed equally to this work.	
    \end{acknowledgements}

\end{document}